\documentclass[10pt,twocolumn,letterpaper]{article}

\usepackage{cvpr}              

\usepackage{graphicx}
\usepackage{amsmath}
\usepackage{amssymb}
\usepackage{booktabs}

\usepackage{algorithm}
\usepackage{algorithmic}
\usepackage{caption}
\usepackage{multirow}
\usepackage{caption}
\usepackage{subcaption}
\usepackage{color}
\usepackage[table,xcdraw]{xcolor}

\usepackage[pagebackref,breaklinks,colorlinks]{hyperref}

\usepackage[capitalize]{cleveref}
\crefname{section}{Sec.}{Secs.}
\Crefname{section}{Section}{Sections}
\Crefname{table}{Table}{Tables}
\crefname{table}{Tab.}{Tabs.}

\def\cvprPaperID{5714} 
\def\confName{CVPR}
\def\confYear{2022}

\begin{document}

\title{Adversarial Attack Based on Prediction-Correction}

\author{Chen Wan, Fangjun Huang\\
School of Computer Science and Engineering, Sun Yat-sen University, Guangzhou, China \\
}
\maketitle

\begin{abstract}
Deep neural networks (DNNs) are vulnerable to adversarial examples obtained by adding small perturbations to original examples. The added perturbations in existing attacks are mainly determined by the gradient of the loss function with respect to the inputs. In this paper, the close relationship between gradient-based attacks and the numerical methods for solving ordinary differential equation (ODE) is studied for the first time. Inspired by the numerical solution of ODE, a new prediction-correction (PC) based adversarial attack is proposed. In our proposed PC-based attack, some existing attack can be selected to produce a predicted example first, and then the predicted example and the current example are combined together to determine the added perturbations. The proposed method possesses good extensibility and can be applied to all available gradient-based attacks easily. Extensive experiments demonstrate that compared with the state-of-the-art gradient-based adversarial attacks, our proposed PC-based attacks have higher attack success rates, and exhibit better transferability.

\end{abstract}

\section{Introduction}
Deep neural networks (DNNs) have achieved remarkable success in numerous areas \cite{1,2,3}. However, recent studies have demonstrated that almost all DNNs are vulnerable to adversarial examples \cite{4,5,6,7,8,9}, which are indistinguishable from original examples but can misguide the DNNs to produce incorrect outputs \cite{10,11,12,13}. The existence of adversarial examples brings many potential security risks to security-sensitive applications, such as face recognition \cite{14} and self-driving automobile \cite{15}. Therefore, the research of adversarial attack is of great significance for enhancing the security and interpretability of DNNs.

Recently, a variety of adversarial attacks have been proposed \cite{9}. Most of the existing adversarial attacks generally fall into the category of white-box attacks, in which the adversary can access all of the information about the target model. As for black-box attacks, the adversary does not know the information of the target model, and sometimes cannot even access the output of the model \cite{16}. Moreover, it is worth noting that the adversarial examples obtained by white-box attacks present cross-model transferability, that is, the adversarial examples crafted for a given model can still fool another model with a high probability \cite{18}, \cite{17}. This property makes it possible to attack DNNs without knowing any information about the models \cite{19}.

According to the way that adversarial examples are generated, the existing attacks can also be divided into three categories, namely, gradient-based attack \cite{10}, optimization-based attack \cite{8, 20}, and generative adversarial networks-based attack \cite{21}. Among these methods, the gradient-based attack has attracted more and more attention because of its lower computational cost and better performance \cite{5,22,23,24}. It is well known that gradient-based attacks utilize the gradients of the loss function with respect to the input example to calculate the adversarial perturbation. In our opinion, there exists some similarity between the gradient-based attacks and the numerical methods for solving ordinary differential equation (ODE). The basic philosophy of the numerical methods is to solve the approximate value of the exact solution through a given initial value and differential coefficients \cite{25, 26}. To be brief, the process of generating adversarial examples in gradient-based attacks can be regarded as the process of solving approximate solutions in the ODE, since they both utilize the gradients (or derivatives) to obtain adversarial examples (or approximate solutions). That is to say, if we regard the gradients in adversarial attacks as the derivatives in the ODE, then the generated adversarial examples can be viewed as the approximate solutions obtained by numerical methods for solving ODE.

So far, although the numerical methods for solving ODE have been widely applied in the field of engineering calculation, there is no research on the relationship between the numerical methods and adversarial attacks, let alone its potential application in the adversarial attacks. In this paper, we make the first attempt in this direction. The close relationship between gradient-based attacks and numerical methods for solving ODE is revealed firstly, and then we propose a series of new prediction-correction (PC) based attacks. In our proposed PC-based attacks, some existing attack is selected to produce a predicted example first, and then the predicted example and the current example are combined together to determine the added perturbations. The main idea of PC-based attacks is to correct the adversarial perturbation through gradient prediction, which is different from the existing attacks that only utilize information from the present and past gradients \cite{10, 22, 27,28,29,30,31}. Compared with the existing gradient-based attacks, our proposed PC-based attacks can achieve higher attack success rates in general. 
The main contributions of this paper can be summarized as follows.
\begin{itemize}
\setlength{\itemsep}{0pt}
\setlength{\parsep}{0pt}
\setlength{\parskip}{0pt}
\setlength{\listparindent}{0em}
\item We first reveal the relationship between the gradient-based attacks and the numerical methods for solving ODE, and establish a close connection between the numerical method and the adversarial attack.
\item A series of new PC-based attacks are proposed in this paper, which can achieve higher attack success rates and significantly improve the transferability.
\item Our proposed prediction-correction strategy possesses good extensibility and can be easily applied to almost all available gradient-based adversarial attacks.
\end{itemize}


\section{Related work}

Let $x$ denote the original image and $y$ denote the corresponding ground-truth label. A given classifier $F\left(x\right)$ outputs a label $\hat{y}=F\left(x\right)$ as the prediction for an input image $x$. The adversarial example $x^{adv}$ is obtained by adding small perturbations to original image $x$ and misclassifies the classifier, \textit{i.e.}, $F\left(x^{adv}\right)\neq y$. The goal of adversarial attack is to maximize the loss function $L\left(x^{adv},y\right)$ of the classifier. In most cases, the $l_p$ norm of the adversarial perturbation is required to be less than a threshold $\epsilon$, \textit{i.e.}, ${\left\| {{x^{adv}} - x} \right\|_p} \le \epsilon $, where $p$ could be $0, 1, 2, \infty$.

\subsection{Gradient-based adversarial attacks}
Among all gradient-based adversarial attacks, FGSM \cite{10} is the most classic one. The existing gradient-based attacks are based on the improvement of FGSM generally. For example, I-FGSM \cite{18}, MI-FGSM \cite{27}, and NI-FGSM \cite{28} are three iterative variants.

Fast Gradient Sign Method (FGSM) \cite{10} generates adversarial example by calculating the gradient of the loss function and performs the one-step update as
\begin{equation}
{x^{adv}} = x + \epsilon  \cdot sign\left( {{\nabla _x}L\left( {x,y} \right)} \right)
\label{eq:1}
\end{equation}
where $\nabla_xL$ is the gradient of the loss function with respect to the original example $x$, and $sign\left(\cdot\right)$ denotes the sign function.

Iterative Fast Gradient Sign Method (I-FGSM) \cite{22} is an iterative attack based on FGSM. This method applies FGSM multiple times with a small step-size $\alpha$ to generate adversarial example as follows.
\begin{equation}
x_{t + 1}^{adv} = clip_x^\epsilon \left\{ {x_t^{adv} + \alpha  \cdot sign\left( {{\nabla _{x_t^{adv}}}L\left( {x_t^{adv},y} \right)} \right)} \right\}
\label{eq:2}
\end{equation}
where $x_0^{adv}=x$, $\alpha  = {\epsilon  \mathord{\left/ {\vphantom {\epsilon  T}} \right. \kern-\nulldelimiterspace} T}$, $T$ represents the number of iterations, and ${clip}_x^\epsilon\left\{\cdot\right\}$ function indications that the generated adversarial example is clipped within the $\epsilon$-ball of the original image.

Momentum Iterative Fast Gradient Sign Method (MI-FGSM) \cite{27} introduces a momentum term into I-FGSM to stabilize the update directions of perturbations, which can be described as follows.
\begin{equation}
{{g}_{t+1}}=\mu \cdot {{g}_{t}}+\frac{{{\nabla }_{x_{t}^{adv}}}L\left( x_{t}^{adv},y \right)}{{{\left\| {{\nabla }_{x_{t}^{adv}}}L\left( x_{t}^{adv},y \right) \right\|}_{1}}}
\label{eq:3}
\end{equation}
\begin{equation}
x_{t+1}^{adv}=clip_{x}^{\epsilon }\left\{ x_{t}^{adv}+\alpha \cdot sign\left( {{g}_{t+1}} \right) \right\}
\label{eq:4}
\end{equation}
where $x_0^{adv}=x$, $g_0=0$, $g_t$ gathers the gradient information up to the $t$-th iteration with a decay factor $\mu$, and $\alpha$ is a step-size.

Nesterov Iterative Fast Gradient Sign Method (NI-FGSM) \cite{28} proposes to improve the transferability of adversarial example by integrating Nesterov accelerated gradient \cite{32} into I-FGSM, which can be formalized as follows.
\begin{equation}
x_{t}^{nes}=x_{t}^{adv}+\alpha \cdot \mu \cdot {{g}_{t}}
\label{eq:5}
\end{equation}
\begin{equation}
{{g}_{t+1}}=\mu \cdot {{g}_{t}}+\frac{{{\nabla }_{x_{t}^{nes}}}L\left( x_{t}^{nes},y \right)}{{{\left\| {{\nabla }_{x_{t}^{nes}}}L\left( x_{t}^{nes},y \right) \right\|}_{1}}}
\label{eq:6}
\end{equation}
\begin{equation}
x_{t+1}^{adv}=clip_{x}^{\epsilon }\left\{ x_{t}^{adv}+\alpha \cdot sign\left( {{g}_{t+1}} \right) \right\}
\label{eq:7}
\end{equation}
where $x_0^{adv}=x$, $g_0=0$, and $x_t^{nes}$ is a state constructed using the previously accumulated gradient information.

\subsection{Data augmentation methods}

The data augmentation method focuses on introducing various input transformations in gradient calculation to prevent the generated adversarial examples from overfitting the model. Introducing data augmentation method into gradient-based attacks (\textit{e.g.}, FGSM \cite{10}, I-FGSM \cite{22}, MI-FGSM \cite{27}, and NI-FGSM \cite{28}) can enhance the transferability of adversarial examples. Several commonly used data augmentation methods are as follows.

Diverse Inputs Method (DIM) \cite{30} first performs random resizing and padding to the input example with a fixed probability, and then feeds the transformed example into the classifier for gradient calculation. 

Translation invariant method (TIM) \cite{31} proposes to use a set of translated images to calculate the gradient. To improve the execution efficiency, the gradient can be approximately calculated by convolving the gradient of the untranslated image with a pre-defined kernel.

Scale Invariant Method (SIM) \cite{28} proposes to calculate the gradient over $m$ scale copies scaled by factor $1/2^i$ on the input example, where $m$ is a hyper-parameter and $i$ ranges from $0$ to $m-1$. 

\section{Methodology}

In this section, we firstly introduce the numerical methods for solving ordinary differential equation (ODE). Then, the close relationship between the numerical methods and gradient-based attacks is uncovered. Based on this, a series of new PC-based attacks are proposed. Finally, the extensibility of the proposed PC-based attacks is discussed in detail.

\subsection{Numerical Methods for Solving ODE}

Considering that the analytical methods are only applicable to some special types of ODE, numerical methods are generally used to solve ODE in practice. Specifically, the numerical methods need to discretize ODE and establish the difference equation, and then the approximate value of the solution (\textit{i.e.}, approximate solution) at some discrete point can be obtained \cite{26}. Let $f\left(t,u\right)$ be a continuous function over the region $G:a\le t\le b$, $\left|u\right|<\infty$. Given an initial value $u_0$, the purpose of numerical methods is to solve the approximate solution $u_n$ satisfying equation (\ref{eq:8}).

\begin{equation}
\left\{ \begin{array}{l}
 \frac{{du}}{{dt}} = f\left( {t,u} \right) \\ 
 u\left( {{t_0}} \right) = {u_0} \\ 
 \end{array} \right.
 \label{eq:8}
\end{equation}

Supposing that the exact solution $u=u\left(t\right)$ of equation (\ref{eq:8}) is unique and smooth enough, the solution region $[a, b]$ can be equally divided into $N$ subintervals with step-size $h$ as described in equation (\ref{eq:9}).
\vspace{-0.1cm}
\begin{equation}
a = {t_0} < {t_1} < {t_2} <  \cdots  < {t_n} <  \cdots  < {t_N} = b
\label{eq:9}
\end{equation}
where $t_n$ represents the node and $t_n=a+nh \ (n=0, \ 1,\cdots,N)$. The numerical method is to find the approximate solution $u_n$ to make sure that it is close to the exact solution $u\left(t\right)$ on node $t_n$  (\textit{i.e.}, $u_n\approx u\left(t_n\right)$). In order to obtain an approximate solution $u_n$, numerical integration is performed on each subinterval $\left[t_n,t_{n+1}\right]$ to establish a difference equation as follows.
\vspace{-0.1cm}
\begin{equation}
u\left( {{t_{n + 1}}} \right) - u\left( {{t_n}} \right) = \int_{{t_n}}^{{t_{n + 1}}} {f\left( {t,u\left( t \right)} \right)dt} 
\label{eq:10}
\end{equation}

Applying the rectangular formula to calculate the integral in equation (\ref{eq:10}), the follow equation can be formed.

\begin{equation}
{u_{n + 1}} = {u_n} + h \cdot f\left( {{t_n},{u_n}} \right)
\label{eq:11}
\end{equation}
where $u_0=u\left(t_0\right)$, and $n=0,\ 1,\cdots, N-1$. According to equation (\ref{eq:11}), the approximate solution $u_{n+1}$ on each node can be obtained one by one. The above method for solving ODE is also called \textbf{Euler method}. However, there generally exists a large error between the approximate solution obtained by Euler method and the exact solution. To reduce the error, the trapezoidal formula can be utilized to calculate the integral on the right side of equation (\ref{eq:10}), and the \textbf{Trapezoidal method} is described as follows.
\vspace{-0.1cm}
\begin{equation}
{u_{n + 1}} = {u_n} + \frac{h}{2} \cdot \left[ {f\left( {{t_n},{u_n}} \right) + f\left( {{t_{n + 1}},{u_{n + 1}}} \right)} \right]
\label{eq:12}
\end{equation}
where $u_0=u\left(t_0\right)$, and $n=0,\ 1,\cdots, N-1$. According to equation (\ref{eq:12}), the approximate solution $u_{n+1}$ on each node can be obtained one by one. Since the Trapezoidal method needs to utilize the value of the future derivative $f\left(t_{n+1},u_{n+1}\right)$ to solve the approximate solution $u_{n+1}$, it is difficult in practical applications. For solving approximate solutions, the \textbf{Improved Euler method} is generally preferred, which is shown below.
\vspace{-0.2cm}
\begin{equation}
u_{n + 1}^{pre} = {u_n} + h \cdot f\left( {{t_n},{u_n}} \right)
\label{eq:13}
\end{equation}
\vspace{-0.4cm}
\begin{equation}
{u_{n + 1}} = {u_n} + \frac{h}{2} \cdot \left[ {f\left( {{t_n},{u_n}} \right) + f\left( {{t_{n + 1}},u_{n + 1}^{pre}} \right)} \right]
\label{eq:14}
\end{equation}
The above equation (\ref{eq:13}) and equation (\ref{eq:14}) are regarded as prediction system and correction system, respectively. Specifically, Euler method is utilized to produce a predicted solution $u_{n+1}^{pre}$, and then fed it into the Trapezoid method for correction to obtain the final approximate solution $u_{n + 1}$. Compared with Euler method, the approximate solution obtained by the Improved Euler method has higher degree of accuracy.

\subsection{Relationship Between Gradient-based Attacks and Numerical Methods for Solving ODE}
It is generally known that I-FGSM \cite{22} is a representative method in gradient-based attacks, and in this place, we take it as an example to study the close relationship between the gradient-based attacks and the numerical methods for solving ODE. According to equation (\ref{eq:2}), neglecting the nonlinear functions, \textit{i.e.}, $clip\left(\cdot\right)$ and $sign\left(\cdot\right)$, I-FGSM can be simplified as 
\begin{equation}
\left\{ \begin{array}{l}
 \frac{{x_{t + 1}^{adv} - x_t^{adv}}}{\alpha } = {\nabla _{x_t^{adv}}}L\left( {x_t^{adv},y} \right) \\ 
 x_0^{adv} = x \\ 
 \end{array} \right.
 \label{eq:15}
\end{equation}
As shown in the above equation, ${\left( {x_{t + 1}^{adv} - x_t^{adv}} \right)}
$ represents the difference between the obtained examples in two adjacent iterations, and $\alpha$ represents the magnitude of the added perturbance at each iteration. If $\alpha \to 0$, then equation (\ref{eq:15}) can be rewritten as follows.
\begin{equation}
\left\{ \begin{array}{l}
 \frac{{dx_t^{adv}}}{{dt}} = {\nabla _{x_t^{adv}}}L\left( {x_t^{adv},y} \right) \\ 
 x_0^{adv} = x \\ 
 \end{array} \right.
 \label{eq:16}
\end{equation}
Specifically, the gradient ${\nabla _{x_t^{adv}}}L\left( {x_t^{adv},y} \right)
$ on the right side of equation (\ref{eq:16}) can be approximately seen as the derivative function $f\left( {t,u} \right)$ in equation (\ref{eq:8}). From this perspective, the process of generating adversarial example can be regarded as the process of seeking approximate solution of ODE. Therefore, it is reasonable to associate gradient-based attacks with numerical methods of ODE since they both utilize the gradients (or derivatives) to obtain the adversarial examples (or approximate solutions).

In addition, we can find that the idea of Euler method for solving ODE has been applied to some existing gradient-based attacks.
It is observed from equations (\ref{eq:1}) and (\ref{eq:11}) that the process of using FGSM to find the adversarial example is almost the same as that of using Euler method to solve ODE. Neglecting the nonlinear functions, \textit{i.e.}, $clip\left(\cdot\right)$ and $sign\left(\cdot\right)$, the equation (\ref{eq:1}) can be described as follows.
\vspace{-0.1cm}
\begin{equation}
{x^{adv}} = x + \epsilon  \cdot {\nabla _x}L\left( {x,y} \right)
\label{eq:17}
\vspace{-0.05cm}
\end{equation}

\noindent By viewing the original example $x$ as the initial value $u_0$ in ODE, and the gradient ${\nabla _x}L\left( {x,y} \right)$ in equation (\ref{eq:17}) as the $f\left( {{t_0},{u _0}} \right)$ in equation (\ref{eq:11}), the adversarial example ${x^{adv}}$ generated by FGSM can be viewed as the approximate solution $u_1$ obtained by Euler method. 

\subsection{Proposed Method}
As mentioned earlier, the adversarial examples generated by FGSM can be regarded as the approximate solutions obtained by Euler method. However, the accuracy of the approximate solution obtained by Euler method is low, and there is a large error between the exact solution and the approximate solution in general. Inspired by the idea of the Improved Euler method, the prediction-correction strategy can be introduced into FGSM \cite{10} to form the prediction-correction based FGSM (PC-FGSM) as follows.
\vspace{-0.05cm}
\begin{equation}
{x^{pre}} = clip_x^\epsilon \left\{ {x + \epsilon  \cdot sign\left( {{\nabla _x}L\left( {x,y} \right)} \right)} \right\}
\label{eq:18}
\end{equation}
\begin{equation}
G = \frac{{{\nabla _x}L\left( {x,y} \right)}}{{{{\left\| {{\nabla _x}L\left( {x,y} \right)} \right\|}_1}}} + \frac{{{\nabla _{{x^{pre}}}}L\left( {{x^{pre}},y} \right)}}{{{{\left\| {{\nabla _{{x^{pre}}}}L\left( {{x^{pre}},y} \right)} \right\|}_1}}}
\label{eq:19}
\end{equation}
\begin{equation}
{x^{adv}} = clip_x^\epsilon \left\{ {x + \epsilon  \cdot sign\left( G \right)} \right\}
\label{eq:20}
\end{equation}
where $\nabla_xL\left(x,y\right)$ and $\nabla_{x^{pre}}L\left(x^{pre},y\right)$ represent the gradients of the loss function with respect to the original example $x$ and predicted example $x^{pre}$, respectively, which are called \textbf{original gradient} and \textbf{predicted gradient} for simplicity in the next. In the above equations, $sign\left(\cdot\right)$ is the sign function to restrict the perturbation in the $l_\infty$ norm bound, and $G$ represents the corrected gradient. As shown in equation (\ref{eq:18}), the adversarial example obtained by FGSM is taken as the predicted example $x^{pre}$, and the normalized predicted gradient is used to correct the normalized original gradient to obtain the corrected gradient $G$. According to the gradient $G$, the perturbation is added to the original example $x$ to generate adversarial example $x^{adv}$.

It should be emphasized that in equations (\ref{eq:18})--(\ref{eq:20}), only one predicted gradient is utilized to correct the added perturbation. In fact, we can iterate the predicted example for many times to obtain multiple predicted gradients, and then use them for correction. As before, the proposed PC-FGSM consists of two steps: prediction and correction. The core process of prediction can be formalized as
\begin{equation}
x_{k + 1}^{pre} = clip_x^\epsilon \left\{ {x_k^{pre} + \epsilon  \cdot sign\left( {{\nabla _{x_k^{pre}}}L\left( {x_k^{pre},y} \right)} \right)} \right\}
\label{eq:21}
\end{equation}
\begin{equation}
G_{k + 1}^{pre} = G_k^{pre} + \frac{{{\nabla _{x_{k + 1}^{pre}}}L\left( {x_{k + 1}^{pre},y} \right)}}{{K \cdot {{\left\| {{\nabla _{x_{k + 1}^{pre}}}L\left( {x_{k + 1}^{pre},y} \right)} \right\|}_1}}}
\label{eq:22}
\end{equation}
where $x_0^{pre}=x$, $G_0^{pre}=0$, $x_k^{pre}$ represents the predicted example obtained when the number of predictions is $k$, $G_k^{pre}$ is the accumulation of the $k$ normalized predicted gradients, and $K$ is the pre-set number of predictions. Please note that the value of $k$ ranges from 0 to $K-1$. In the algorithm, as shown in equations (\ref{eq:21}) and (\ref{eq:22}), $K$ predicted examples (\textit{i.e.}, $x_1^{pre}, \ x_2^{pre}, \cdots, \ x_K^{pre}$) are computed firstly, and then the average of the normalized gradients of these predicted examples is calculated to obtain $G_K^{pre}$. 

The core process of correction can be formalized as follows.
\begin{equation}
G = \frac{{{\nabla _x}L\left( {x,y} \right)}}{{{{\left\| {{\nabla _x}L\left( {x,y} \right)} \right\|}_1}}} + G_K^{pre}
\label{eq:23}
\end{equation}
\begin{equation}
{x^{adv}} = clip_x^\epsilon \left\{ {x + \epsilon  \cdot sign\left( G \right)} \right\}
\label{eq:24}
\end{equation}
where $G_K^{pre}$ represents the average of the normalized predicted gradients. As shown in equation (\ref{eq:23}), $G_K^{pre}$ is combined with the normalized original gradient to obtain the corrected gradient $G$ firstly. Then, according to the corrected gradient $G$, the adversarial example $x^{adv}$ can be generated by adding the perturbation to the original example $x$.

\begin{algorithm}[t]
\caption{PC-FGSM}
\label{alg:algorithm1}
\textbf{Input}: Original example $x$ with ground-truth label $y$, a classifier $F$ with loss function $L$.\\
\textbf{Parameter}: The magnitude of perturbation $\epsilon$, number of predictions $K$.\\
\textbf{Output}: Adversarial example $x^{adv}$.

\begin{algorithmic}[1] 
\STATE $x_0^{pre}=x$, $G_0^{pre}=0$
\STATE Input $x$ to $F$ and obtain the original gradient $\nabla_{x}L\left(x,y\right)$

\FOR{$k=0$ to $K-1$} 
\STATE Update $x_{k+1}^{pre}$ by equation (\ref{eq:21})
\STATE Calculate the gradient $\nabla_{x_{k+1}^{pre}}L\left(x_{k+1}^{pre},y\right)$
\STATE Update $G_{k + 1}^{pre}$ by equation (\ref{eq:22})
\ENDFOR
\STATE Obtain the corrected gradient $G$ by equation (\ref{eq:23})
\STATE Obtain the adversarial example $x^{adv}$ by equation (\ref{eq:24})
\STATE \textbf{return} $x^{adv}$.
\end{algorithmic}
\end{algorithm}

The complete PC-FGSM is summarized in Algorithm \ref{alg:algorithm1}. Note that in our proposed PC-FGSM, the number of gradient calculations is $K+1$ (\textit{i.e.},  $K$ predicted gradients and one original gradient), where $K$ represents the number of predictions.

\subsection{Extensibility of the PC-based Attack}
The idea of prediction-correction can be naturally introduced into other FGSM based adversarial attacks to form a series of new prediction-correction based attacks. For example, combined with I-FGSM \cite{22}, we can obtain PC-I-FGSM.

\begin{figure}[h]
\vspace{-0.35cm}
\centering
\includegraphics[width=0.68\columnwidth]{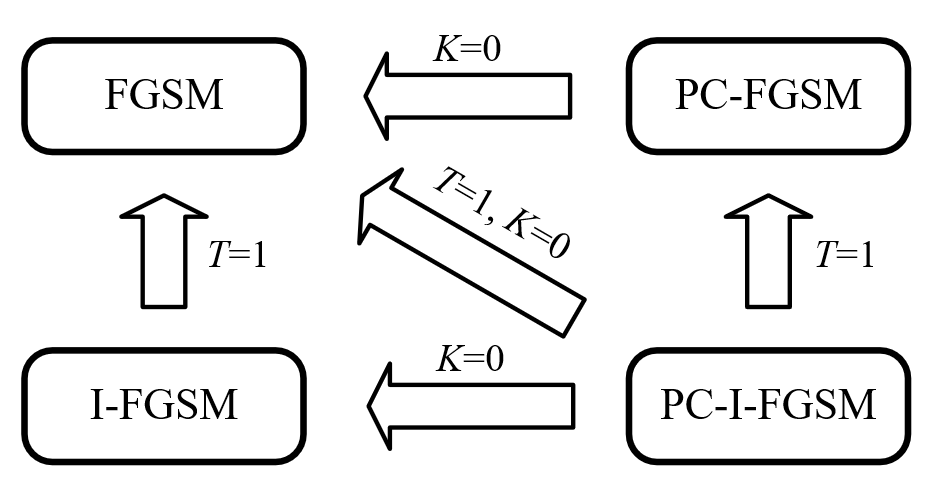}
\centering
\caption{Relationships between different attacks.}
\label{Fig:1}
\vspace{-0.1cm}
\end{figure}

The above-mentioned attacks (\textit{i.e.}, FGSM, I-FGSM, PC-FGSM, and PC-I-FGSM) all belong to gradient-based attacks, and the relationships between them are shown in Fig. \ref{Fig:1}. 
\vspace{-0.2cm}
\begin{itemize}
\setlength{\itemsep}{0pt}
\setlength{\parsep}{0pt}
\setlength{\parskip}{0pt}
\setlength{\listparindent}{0em}
\item If the number of predictions $K=0$, PC-FGSM degrades to FGSM, and PC-I-FGSM degrades to I-FGSM. 
\item If the number of iterations $T=1$, I-FGSM degrades to FGSM, and PC-I-FGSM degrades to PC-FGSM.  
\item If the number of predictions $K=0$ and iterations $T=1$, PC-I-FGSM degrades to FGSM.
\vspace{-0.2cm}
\end{itemize}

In addition, the PC strategy proposed in this paper can be combined with MI-FGSM \cite{27} and NI-FGSM \cite{28} to form PC-MI-FGSM and PC-NI-FGSM, respectively. Meanwhile, we can introduce the data augmentation methods (\textit{e.g.}, DIM \cite{30}, TIM \cite{31}, and SIM \cite{28}) into PC-based attacks to improve the transferability of adversarial examples. Please note that in our algorithm, in each iteration, there are $K+1$ examples (\textit{i.e.}, $K$ predicted examples and one current example). For each example, the corresponding loss function can be obtained, and the gradients are computed according to the obtained loss function. The data augmentation methods can be applied to these examples directly. For example, PC-NI-FGSM can be integrated with DIM \cite{30}, TIM \cite{31}, and SIM \cite{28} to form a series of new powerful adversarial attacks, named PC-DI-NI-FGSM, PC-TI-DI-NI-FGSM, and PC-SI-TI-DI-NI-FGSM. The detailed algorithms for these attacks are provided in Appendix A of the attached supplementary material.

\section{Experiments}
\subsection{Experiment Settings}
All the experiments are performed on the TensorFlow DNN computing framework \cite{33} and run with four parallel NVIDIA GeForce GTX 1080Ti GPUs. 
We use the $l_\infty$ norm to measure the distortion, and the cross-entropy loss function is selected in our experiments.
The dataset, models, baselines, and hyper-parameters are set as follows.

\noindent \textbf{Dataset}. The test dataset consists of 10, 000 images randomly selected from the ImageNet validation set \cite{34}, and almost all can be correctly classified by the model exploited in this paper. All these images are resized to $299\times299\times3$ beforehand.

\noindent \textbf{Models}. In the experiment, 14 models are selected for testing, including four normally trained models and ten defense models. 
\vspace{-0.18cm}
\begin{itemize}
\setlength{\itemsep}{0pt}
\setlength{\parsep}{0pt}
\setlength{\parskip}{0pt}
\setlength{\listparindent}{0em}
\item Four normally trained models, \textit{i.e.}, Inception-v3 (Inc-v3) \cite{35}, Inception-v4 (Inc-v4) \cite{36}, Inception-Resnet-v2 (IncRes-v2) \cite{36} and Resnet-v2-152 (Res-152) \cite{37}. 
\item Ten defense models
\vspace{-0.15cm}
\begin{itemize}
\setlength{\itemsep}{0pt}
\setlength{\parsep}{0pt}
\setlength{\parskip}{0pt}
\setlength{\listparindent}{0em}
\item three adversarially trained models, \textit{i.e.}, Inc-v3{\tiny ens3}, Inc-v3{\tiny ens4}, and IncRes-v2{\tiny ens} \cite{38}); 
\item top-3 defense models in NIPS 2017 competition, \textit{i.e.}, high-level representation guided denoiser (HGD rank-1) \cite{39}, random resizing and padding (R{\rm{\&}}P rank-2) \cite{40} and the rank-3 submission (NIPS-r3\footnote{https://github.com/anlthms/nips-2017/tree/master/mmd}); 
\item four recently proposed defense methods, \textit{i.e.}, purifying perturbations via image compression model (Comdefend) \cite{41}, randomized smoothing (RS) \cite{42}, feature distillation (FD$_1$) \cite{43}, and feature denoising (FD$_2$) \cite{44}.
\end{itemize}
\end{itemize}

\noindent \textbf{Baselines}. Four gradient-based adversarial attacks (\textit{i.e.}, FGSM \cite{10}, I-FGSM \cite{22}, MI-FGSM \cite{27}, and NI-FGSM \cite{28}) are selected as the baselines to compare with our proposed PC-based attacks, including PC-FGSM, PC-I-FGSM, PC-MI-FGSM, and PC-NI-FGSM. In addition, three data augmentation methods (\textit{i.e.}, DIM \cite{30}, TIM \cite{31}, and SIM \cite{28}) are integrated into NI-FGSM \cite{28} and our proposed PC-NI-FGSM for comparison. NI-FGSM integrated with the three augmentation methods are named DI-NI-FGSM, TI-DI-NI-FGSM, and SI-TI-DI-NI-FGSM, and PC-NI-FGSM integrated with the three data augmentation methods are termed PC-DI-NI-FGSM, PC-DI-NI-FGSM, and PC-SI-TI-DI-NI-FGSM, respectively.

\noindent \textbf{Hyper-parameters}. The maximum perturbation, number of iterations, number of predictions, step-size, and decay factor are set as $\epsilon=16$, $T=10$, $K=1$, $\alpha  = {\epsilon  \mathord{\left/
 {\vphantom {\epsilon  T}} \right.
 \kern-\nulldelimiterspace} T} = 1.6
$, and $\mu=1.0$. For three augmentation methods, the transformation probability of DIM \cite{30} is set to $0.5$, the Gaussian kernel with size $15\times15$ is selected as the kernel of TIM \cite{31}, and the number of scale copies of SIM \cite{28} is $5$.

\begin{table*}[t]\footnotesize
\begin{center}
\begin{tabular}{llccccccc}
\toprule
Model                                       & Attack         & Inc-v3          & Inc-v4          & IncRes-v2       & Res-152         & Inc-v3{\tiny ens3}     & Inc-v3{\tiny ens4}     & IncRes-v2{\tiny ens}  \\ \midrule
\multicolumn{1}{c}{\multirow{2}{*}{Inc-v3}} & FGSM           & 65.17$^*$          & 27.17           & 25.87           & 25.52           & 9.12           & 8.53           & \textbf{4.03} \\
\multicolumn{1}{c}{}                        & PC-FGSM (Ours) & \textbf{94.76}$^*$ & \textbf{49.17}  & \textbf{47.58}  & \textbf{39.22}  & \textbf{9.80}  & \textbf{8.89}  & 3.77          \\ \midrule
\multicolumn{1}{c}{\multirow{2}{*}{Inc-v4}}                     & FGSM           & 30.65           & 51.49$^*$          & 23.74           & 24.80           & 9.96           & 8.93           & 4.56          \\
                                            & PC-FGSM (Ours) & \textbf{58.31}  & \textbf{87.42}$^*$ & \textbf{49.59}  & \textbf{43.39}  & \textbf{11.54} & \textbf{9.75}  & \textbf{4.63} \\ \midrule
\multicolumn{1}{c}{\multirow{2}{*}{IncRes-v2}}                  & FGSM           & 28.06           & 22.67           & 41.06$^*$          & 22.78           & 10.14          & 9.10           & 5.44          \\
                                            & PC-FGSM (Ours) & \textbf{52.87}  & \textbf{47.70}  & \textbf{76.48}$^*$ & \textbf{41.67}  & \textbf{13.14} & \textbf{10.79} & \textbf{6.36} \\ \midrule
\multicolumn{1}{c}{\multirow{2}{*}{Res-152}}                   & FGSM           & 37.42           & 31.59           & 30.69           & 75.00$^*$          & 14.62          & 12.18          & 6.81          \\
                                            & PC-FGSM (Ours) & \textbf{58.69}  & \textbf{53.54}  & \textbf{52.47}  & \textbf{97.17}$^*$ & \textbf{16.29} & \textbf{13.13} & \textbf{6.82} \\ \bottomrule
\end{tabular}
\vspace{-0.15cm}
\caption{Attack success rates ($\%$) of the adversarial examples against the seven models, $^*$ indications the white-box attacks.}
\label{Tab:1}
\end{center}
\vspace{-0.2cm}
\end{table*}

\begin{table*}[t]\footnotesize
\begin{center}
\vspace{-0.15cm}
\begin{tabular}{lccccccc}
\toprule
\multicolumn{1}{c}{Attack}   & Inc-v3          & Inc-v4         & IncRes-v2      & Res-152        & Inc-v3{\tiny ens3}     & Inc-v3{\tiny ens4}     & IncRes-v2{\tiny ens}   \\ \midrule
I-FGSM                       & 99.91$^*$         & 24.66          & 19.60          & 14.76          & 4.93           & 3.85           & 2.35           \\
PC-I-FGSM (Ours)             & \textbf{99.92}$^*$ & \textbf{27.63} & \textbf{22.38} & \textbf{16.76} & \textbf{5.18}  & \textbf{4.35}  & \textbf{2.61}  \\ \midrule
MI-FGSM                      & \textbf{99.89}$^*$          & 47.87          & 43.55          & 36.04          & \textbf{13.07} & \textbf{12.11} & \textbf{6.27}  \\
PC-MI-FGSM (Ours)            & \textbf{99.89}$^*$ & \textbf{51.89} & \textbf{48.03} & \textbf{37.67} & 12.99          & 11.64          & 6.12           \\ \midrule
NI-FGSM                      & 99.90$^*$          & 55.02          & 51.15          & 39.71          & 13.02          & 11.58          & 5.94           \\
PC-NI-FGSM (Ours)            & \textbf{99.93}$^*$ & \textbf{60.65} & \textbf{56.20} & \textbf{44.22} & \textbf{14.13} & \textbf{12.44} & \textbf{6.28}  \\ \midrule
DI-NI-FGSM                   & 99.90$^*$          & 64.53          & 59.81          & 46.58          & 14.10          & 13.04          & 6.82           \\
PC-DI-NI-FGSM (Ours)         & \textbf{99.93}$^*$ & \textbf{79.60} & \textbf{75.09} & \textbf{61.09} & \textbf{18.62} & \textbf{17.70} & \textbf{8.86}  \\ \midrule
TI-DI-NI-FGSM                & 99.61$^*$          & 56.52          & 46.14          & 41.06          & 38.11          & 36.20          & 27.19          \\
PC-TI-DI-NI-FGSM (Ours)    & \textbf{99.84}$^*$ & \textbf{67.58} & \textbf{58.04} & \textbf{50.90} & \textbf{48.11} & \textbf{47.14} & \textbf{36.98} \\ \midrule
SI-TI-DI-NI-FGSM             & \textbf{99.94}$^*$ & 73.94          & 64.49          & 58.63          & 60.53          & 59.99          & 47.56          \\
PC-SI-TI-DI-NI-FGSM (Ours) & 99.90$^*$          & \textbf{81.44} & \textbf{72.45} & \textbf{66.70} & \textbf{70.63} & \textbf{70.41} & \textbf{57.41} \\ \bottomrule
\end{tabular}
\vspace{-0.15cm}
\caption{Attack success rates ($\%$) of the adversarial examples obtained by Inc-v3 against the seven models, $^*$ indications the white-box attacks.}
\label{Tab:2}
\end{center}
\vspace{-0.66cm}
\end{table*}

\subsection{Attacking the Single model}
Firstly, we separately use PC-FGSM and FGSM to attack four normally trained models (\textit{i.e.}, Inc-v3, Inc-v4, IncRes-v2, and Res-152) to obtain eight groups of adversarial examples. Then, the obtained adversarial examples are tested on seven models (\textit{i.e.}, Inc-v3, Inc-v4, IncRes-v2, Res-152, Inc-v3{\tiny ens3}, Inc-v3{\tiny ens4}, and IncRes-v2{\tiny ens}). The attack success rates of the obtained adversarial examples against seven models are shown in Table \ref{Tab:1}. It can be observed that the attack success rates of PC-FGSM are $22.17\% \sim 35.93\%$ higher than that of FGSM in white-box models, $13.70 \% \sim 27.66\%$ higher than that of FGSM in black-box models, and $0.01\% \sim 3.00\%$ higher than that of FGSM in defense models. The experimental results demonstrate that our proposed PC-FGSM can achieve higher attack success rates, and possess better transferability.

Next, the PC strategy is introduced into the existing iterative attacks to evaluate the effectiveness of our method. As mentioned above, twelve gradient-based iterative attacks are selected for comparison, including six baseline attacks (\textit{i.e.}, I-FGSM, MI-FGSM, NI-FGSM, DI-NI-FGSM, TI-DI-NI-FGSM, and SI-TI-DI-NI-FGSM) and the corresponding six PC-based attacks (\textit{i.e.}, PC-I-FGSM, PC-MI-FGSM, PC-NI-FGSM, PC-DI-NI-FGSM, PC-TI-DI-NI-FGSM, and PC-SI-TI-DI-NI-FGSM). These twelve attacks are used to attack four normally trained models (\textit{i.e.}, Inc-v3, Inc-v4, IncRes-v2, and Res-152) to obtain the adversarial examples firstly. Then, the obtained adversarial examples are tested on seven models, including Inc-v3, Inc-v4, IncRes-v2, Res-152, Inc-v3{\tiny ens3}, Inc-v3{\tiny ens4}, and IncRes-v2{\tiny ens}. The attack success rates of adversarial examples obtained by Inc-v3 are reported in Table \ref{Tab:2}.
The results for adversarial examples obtained by other three models are provided in Appendix B of the supplementary material.
As seen, in white-box models the attack success rates of our proposed methods are close to $100\%$, in black-box models the attack success rates of our methods are $1.19\% \sim 17.26\%$ higher than that of the baseline attacks, and in defense models the attack success rates of our methods are $0.04\% \sim 15.24\%$ higher than that of the baseline attacks.

\begin{table*}[!]\footnotesize
\begin{center}
\begin{tabular}{lcccccccc}
\toprule
\multicolumn{1}{c}{Attack}   & HGD            & R{\rm{\&}}P           & NIPS-r3        & ComDefend      & RS             & FD$_1$            & FD$_2$            & Average        \\ \midrule
FGSM                         & \textbf{0.90}           & 7.67           & 12.02          & 26.97          & 23.43          & 26.20          & 15.95          & 16.16          \\
PC-FGSM (Ours)               & 0.76  & \textbf{10.35} & \textbf{20.81} & \textbf{47.60} & \textbf{30.64} & \textbf{46.88} & \textbf{16.90} & \textbf{24.85} \\ \midrule
I-FGSM                       & 21.55          & 8.95           & 12.38          & 17.22          & 14.04          & 16.43          & 14.89          & 15.07          \\
PC-I-FGSM (Ours)             & \textbf{24.72} & \textbf{9.58}  & \textbf{13.81} & \textbf{17.74} & \textbf{14.22} & \textbf{17.04} & \textbf{14.94} & \textbf{16.01} \\ \midrule
MI-FGSM                      & 32.89          & 24.99          & 33.58          & 43.94          & 23.38          & 43.00          & \textbf{15.89}          & 31.10          \\
PC-MI-FGSM (Ours)            & \textbf{36.23} & \textbf{25.23} & \textbf{35.53} & \textbf{45.34} & \textbf{23.47} & \textbf{45.15} & 15.88 & \textbf{32.40} \\ \midrule
NI-FGSM                      & 26.69          & 22.83          & 31.85          & 42.76          & 23.47          & 42.77          & 15.83          & 29.46          \\
PC-NI-FGSM (Ours)            & \textbf{41.01} & \textbf{30.50} & \textbf{43.24} & \textbf{54.84} & \textbf{25.49} & \textbf{54.13} & \textbf{16.19} & \textbf{37.91} \\ \midrule
DI-NI-FGSM                   & 30.39          & 26.02          & 37.23          & 46.30          & 24.99          & 45.92          & 16.02          & 32.41          \\
PC-DI-NI-FGSM (Ours)         & \textbf{56.06} & \textbf{44.53} & \textbf{58.88} & \textbf{65.67} & \textbf{30.44} & \textbf{64.58} & \textbf{16.63} & \textbf{48.11} \\ \midrule
TI-DI-NI-FGSM                & 75.01 & 68.62          & 70.54          & 70.09          & 55.63          & 71.62          & 18.51          & 61.43          \\
PC-TI-DI-NI-FGSM   (Ours)    & \textbf{86.17}          & \textbf{81.26} & \textbf{82.62} & \textbf{80.75} & \textbf{65.29} & \textbf{82.72} & \textbf{19.69} & \textbf{71.21} \\ \midrule
SI-TI-DI-NI-FGSM             & 90.68          & 86.96          & 88.50          & 88.11          & 75.16          & 89.59          & 22.30          & 77.33          \\
PC-SI-TI-DI-NI-FGSM   (Ours) & \textbf{93.39} & \textbf{90.87} & \textbf{92.04} & \textbf{92.57} & \textbf{82.17} & \textbf{93.49} & \textbf{23.73} & \textbf{81.18} \\ \bottomrule  
\end{tabular}
\vspace{-0.17cm}
\caption{Attack success rates ($\%$) of the adversarial examples (crafted by the ensemble of models) against the seven advance models.}
\label{Tab:3}
\end{center}
\vspace{-0.5cm}
\end{table*}

\subsection{Attacking the Ensemble of Models}
Following the ensemble method described in Dong et al. \cite{27}, we separately use the above-mentioned fourteen methods to attack the ensemble of models (which is composed of Inc-v3, Inc-v4, IncRes-v2, and Res-152 with the same ensemble weight), and fourteen groups of adversarial examples can be obtained. The attack success rates of the obtained fourteen groups of adversarial examples against 
the seven advanced defense models (\textit{i.e.}, HGD, R{\rm{\&}}P, NIPS-r3, ComDefend, RS, FD$_1$, and FD$_2$) are shown in Table \ref{Tab:3}. As seen, the attack success rates of our proposed PC-based attacks against the advanced defense models are $0.15\% \sim 25.67\%$ higher than that of the corresponding baseline attacks in general.

\subsection{Discussion}
In this subsection, the influence of the number of predictions on the attack success rate is discussed firstly. Then, the attack success rates of our PC-based attacks and baseline attacks are compared with the same number of iterations and gradient calculations. 

\vspace{0.12cm}
\noindent\textit{1) Discussion about the number of predictions}
\vspace{0.05cm}

\begin{figure}[!]
    \centering
   \begin{subfigure}[b]{0.22\textwidth}
        \centering
       \includegraphics[width=0.95\textwidth]{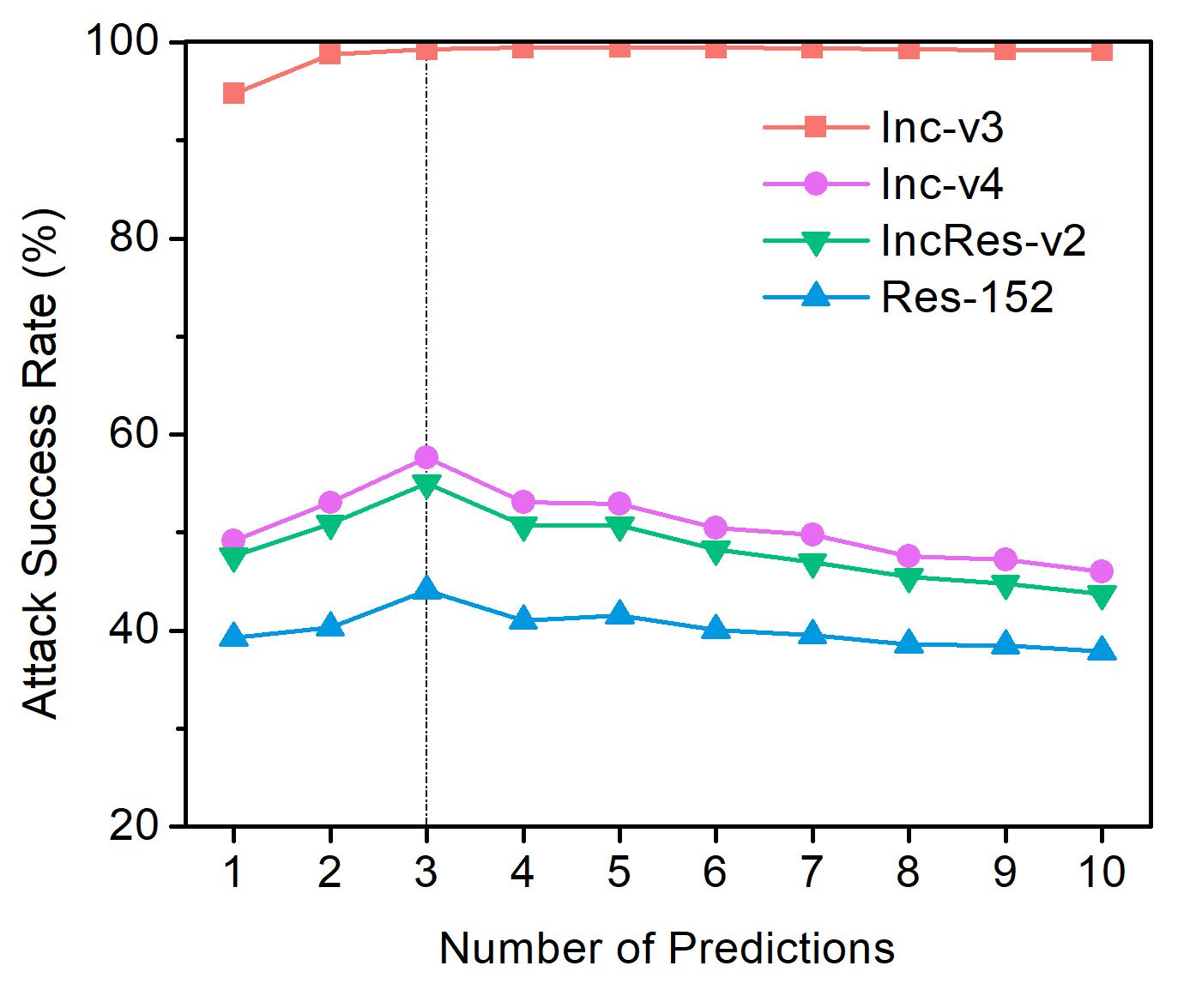}
        \caption{}
        \label{subfigure 2.1}
    \end{subfigure}
     \hfill
   \begin{subfigure}[b]{0.22\textwidth}
        \centering
       \includegraphics[width=0.95\textwidth]{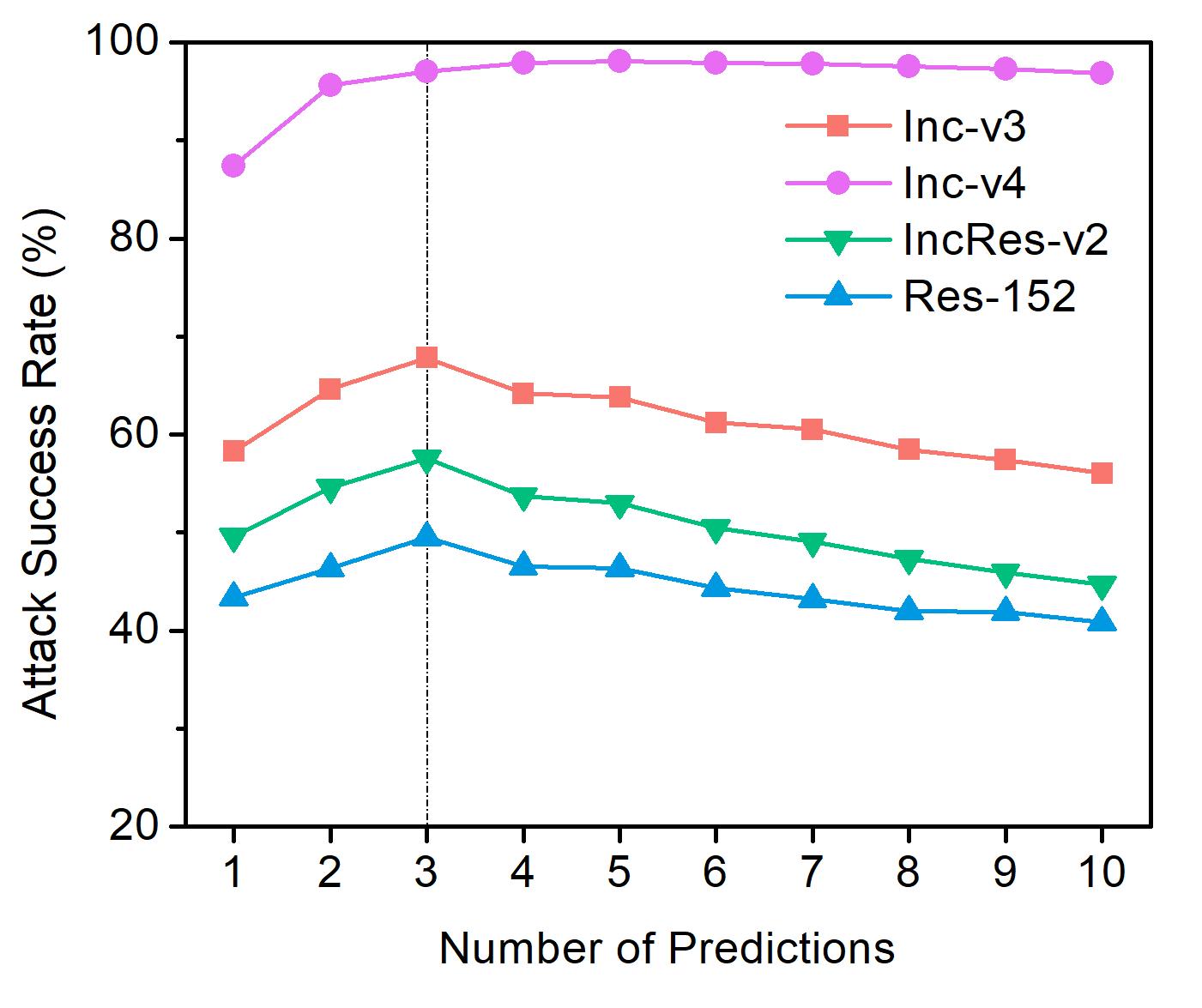}
        \caption{}
        \label{subfigure 2.2}
    \end{subfigure}
    \\
    \begin{subfigure}[b]{0.22\textwidth}
        \centering
       \includegraphics[width=0.95\textwidth]{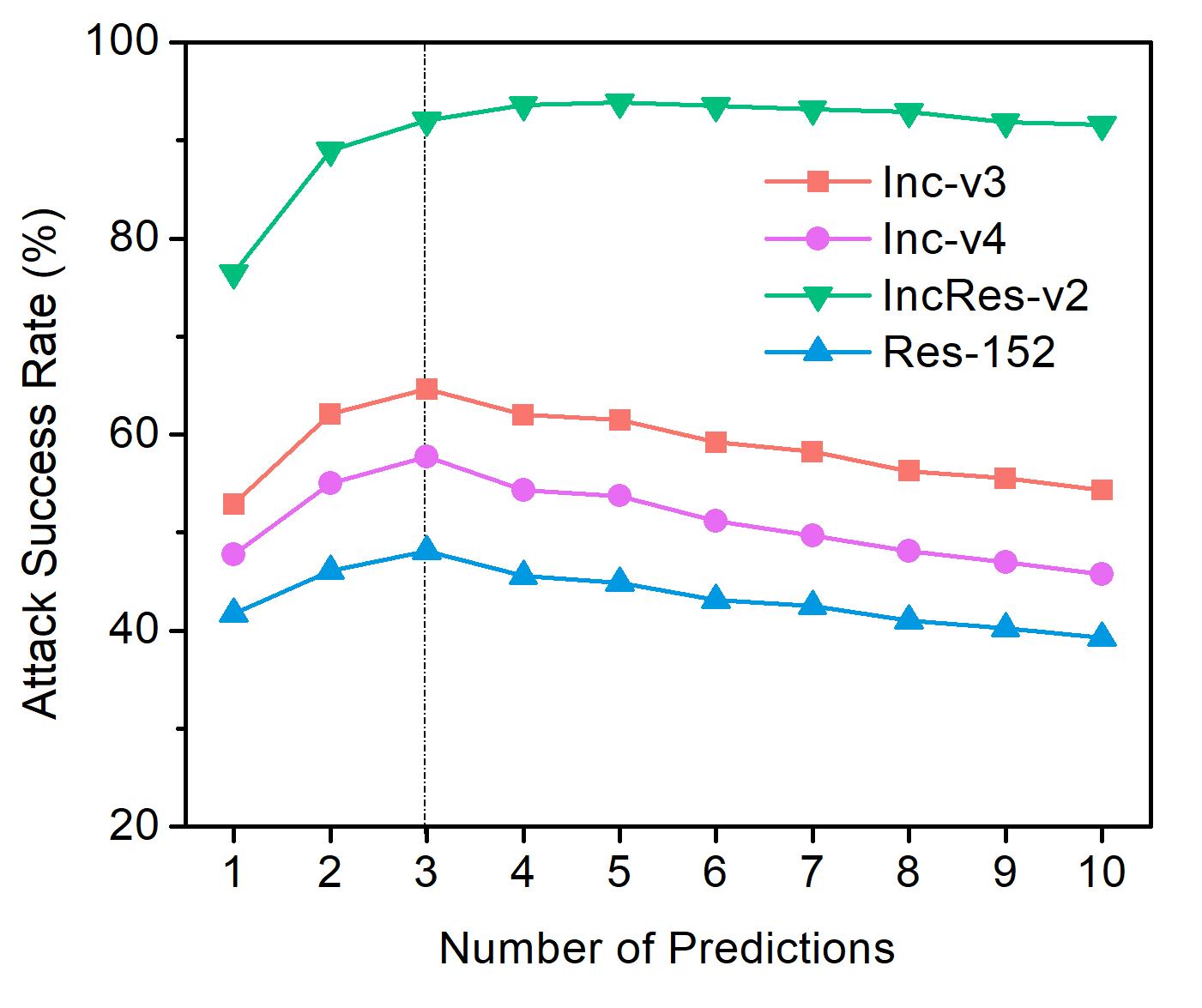}
        \caption{}
        \label{subfigure 2.3}
    \end{subfigure}
        \hfill
    \begin{subfigure}[b]{0.22\textwidth}
        \centering
       \includegraphics[width=0.95\textwidth]{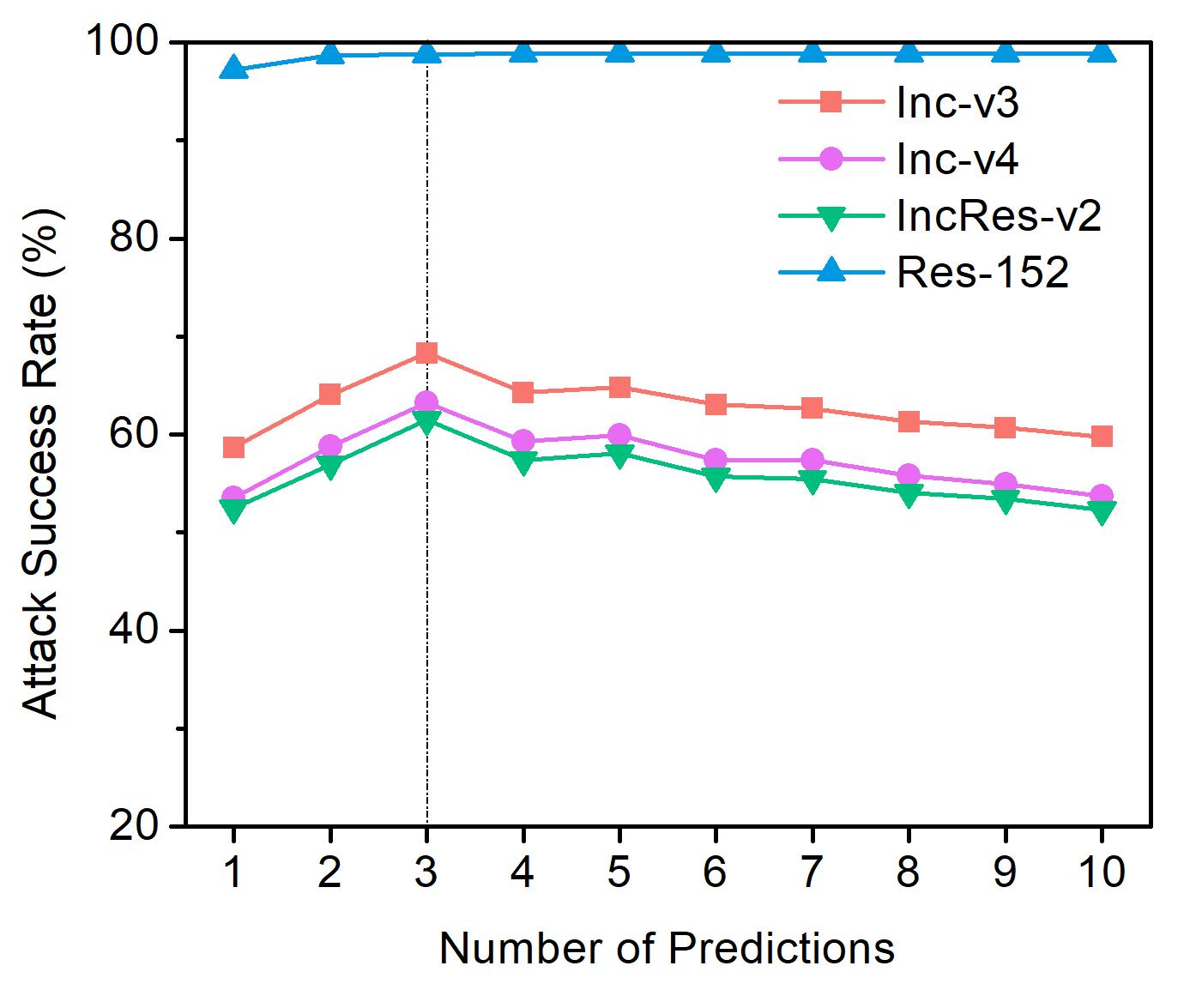}
        \caption{}
        \label{subfigure 2.4}
    \end{subfigure}
    \vspace{-0.2cm} 
\caption{The attack success rates (\%) of the adversarial examples obtained by PC-FGSM with different number of predictions against four normally trained models. (a) The adversarial examples generated via attacking Inc-v3; (b) The adversarial examples generated via attacking Inc-v4; (c) The adversarial examples generated via attacking IncRes-v2; (d) The adversarial examples generated via attacking Res-152.
}
\label{Fig:2}  
\vspace{-0.45cm} 
\end{figure}

Firstly, we utilize the proposed PC-FGSM to attack four normally trained models (\textit{i.e.}, Inc-v3, Inc-v4, IncRes-v2, and Res-152) to generate adversarial examples, where the number of predictions $K$ is selected in the range of 1 to 10 and the maximum perturbation is set as $\epsilon=16$. Then, the obtained adversarial examples are tested on the aforementioned four models. The experimental results are shown in Figure \ref{Fig:2}, where the abscissa represents the number of predictions and the ordinate represents the attack success rate. It can be observed that with the increase of the number of predictions, the attack success rate increases at first and then decreases. When the number of predictions is about 3, our proposed PC-FGSM generally can achieve the highest attack success rate.

Next, we further discuss the impact of the number of predictions on our proposed PC-based iterative attacks (\textit{i.e.}, PC-I-FGSM, PC-MI-FGSM, and PC-NI-FGSM). 
For detailed experiments, please refer to Appendix C in the supplementary material. The experimental results demonstrate that
for these iterative attacks, increasing the number of predictions $K$ does not necessarily improve the attack success rates of our PC-based attacks.
Considering that increasing the number of predictions will increase the computational complexity of our algorithm, the number of predictions $K$ is generally set to 1 in practical applications for those iterative attacks.

\vspace{0.12cm}
\noindent\textit{2) With the same number of iterations}
\vspace{0.05cm}

In this part, three PC-based iterative attacks (\textit{i.e.}, PC-I-FGSM, PC-MI-FGSM, and PC-NI-FGSM) and three baseline iterative attacks (\textit{i.e.}, I-FGSM, MI-FGSM, and NI-FGSM) are selected for comparison. In the experiment, the number of predictions, maximum perturbation, and step-size are set as $K=1$, $\epsilon=16$, and $ \alpha  = {\epsilon  \mathord{\left/ {\vphantom {\epsilon  T}} \right. \kern-\nulldelimiterspace} T} $. The above six methods are used to attack Inc-v3 to obtain adversarial examples, where the number of iterations $T$ is selected in the range of 1 to 10. 
The attack success rates of the generated adversarial examples against four normally trained models are shown in Figure \ref{Fig:3}, and the experimental results of NI-FGSM and PC-NI-FGSM are shown in Appendix C. Please note that in Figure \ref{Fig:3}, the label “model vs. method” represents the attack success rates of the adversarial examples (generated by “method”) against “model”.
As seen, with the same number of iterations, our proposed PC-based attacks consistently achieve higher success rates than those baseline attacks.

\begin{figure}[hbt!]
    \centering
   \begin{subfigure}[b]{0.22\textwidth}
        \centering
       \includegraphics[width=0.95\textwidth]{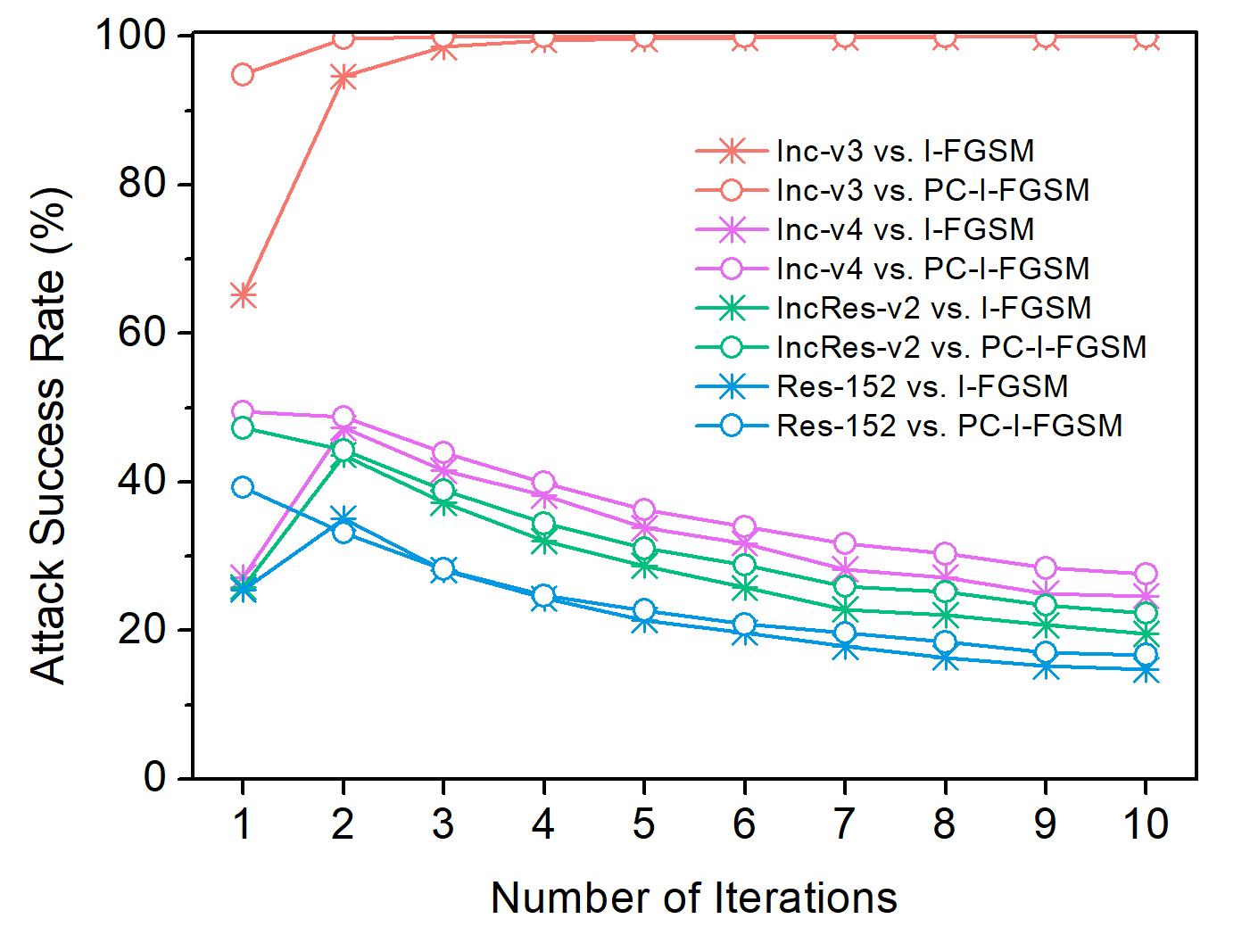}
        \caption{}
        \label{subfigure 3.1}
    \end{subfigure}
     \hfill
   \begin{subfigure}[b]{0.22\textwidth}
        \centering
       \includegraphics[width=0.95\textwidth]{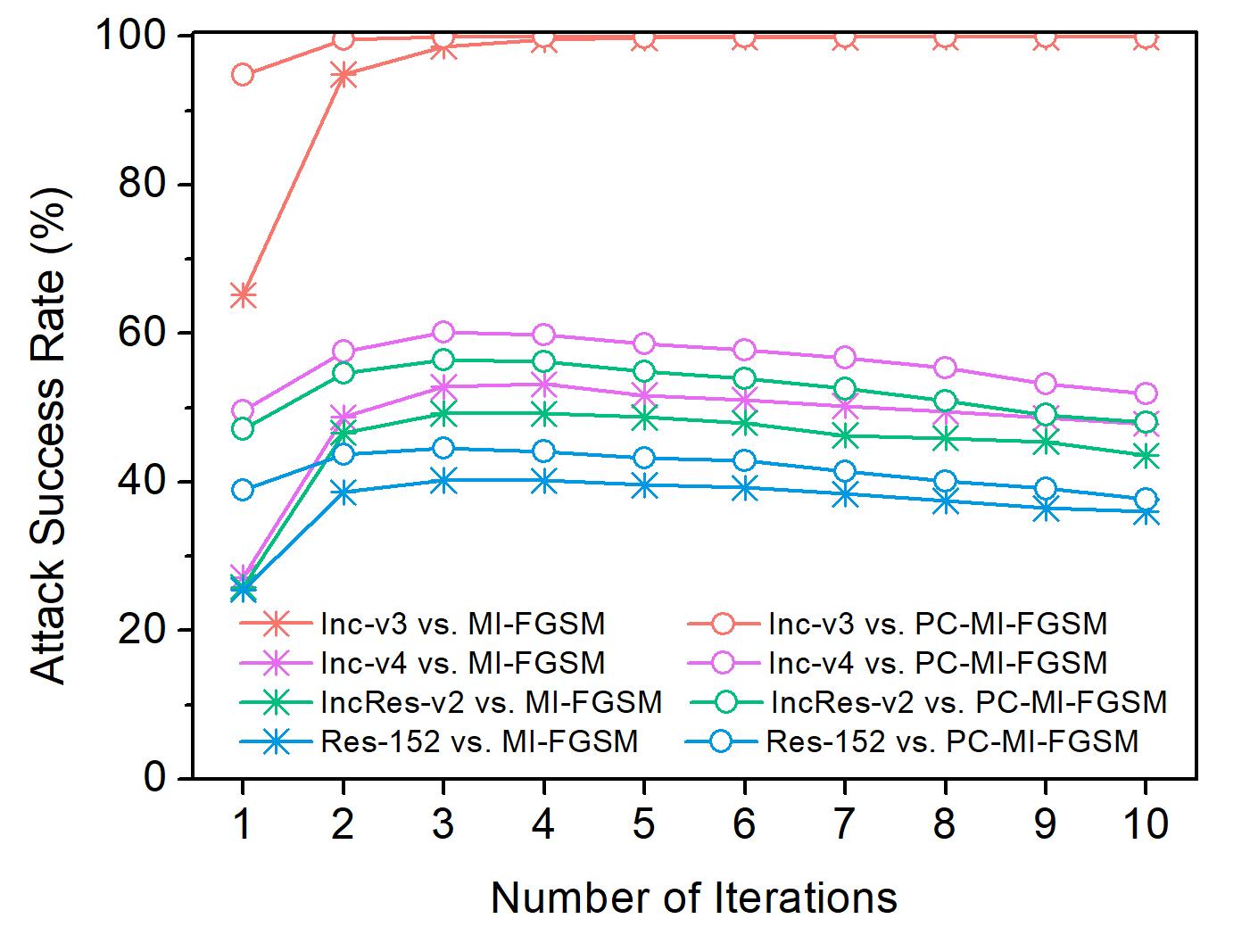}
        \caption{}
        \label{subfigure 3.2}
    \end{subfigure}
\vspace{-0.3cm}
\caption{The attack success rates (\%) of our proposed PC-based attacks and the baseline attacks with the same number of iterations. (a) Comparison of I-FGSM and PC-I-FGSM; (b) Comparison of MI-FGSM and PC-MI-FGSM.}
\label{Fig:3}   
\vspace{-0.1cm}
\end{figure}

\noindent\textit{3) With the same number of gradient calculations}
\vspace{0.05cm}

In this part, the attack success rates of our proposed PC-based attacks and three baseline attacks (\textit{i.e.}, I-FGSM, MI-FGSM, and NI-FGSM) are compared with the same number of gradient calculations. 
Note that for our proposed PC-based attacks, the number of gradient calculations is $T\times(K+1)$, where $T$ and $K$ represent the number of iterations and predictions, respectively. For the three baseline attacks, the number of gradient calculations is equal to the number of iterations $T$.
In the experiment, the maximum perturbation and the step-size are set as $\epsilon=16$ and $ \alpha  = {\epsilon  \mathord{\left/ {\vphantom {\epsilon  T}} \right. \kern-\nulldelimiterspace} T}$.

\begin{figure}[hbt!]
    \centering
   \begin{subfigure}[b]{0.22\textwidth}
        \centering
       \includegraphics[width=1.1\textwidth]{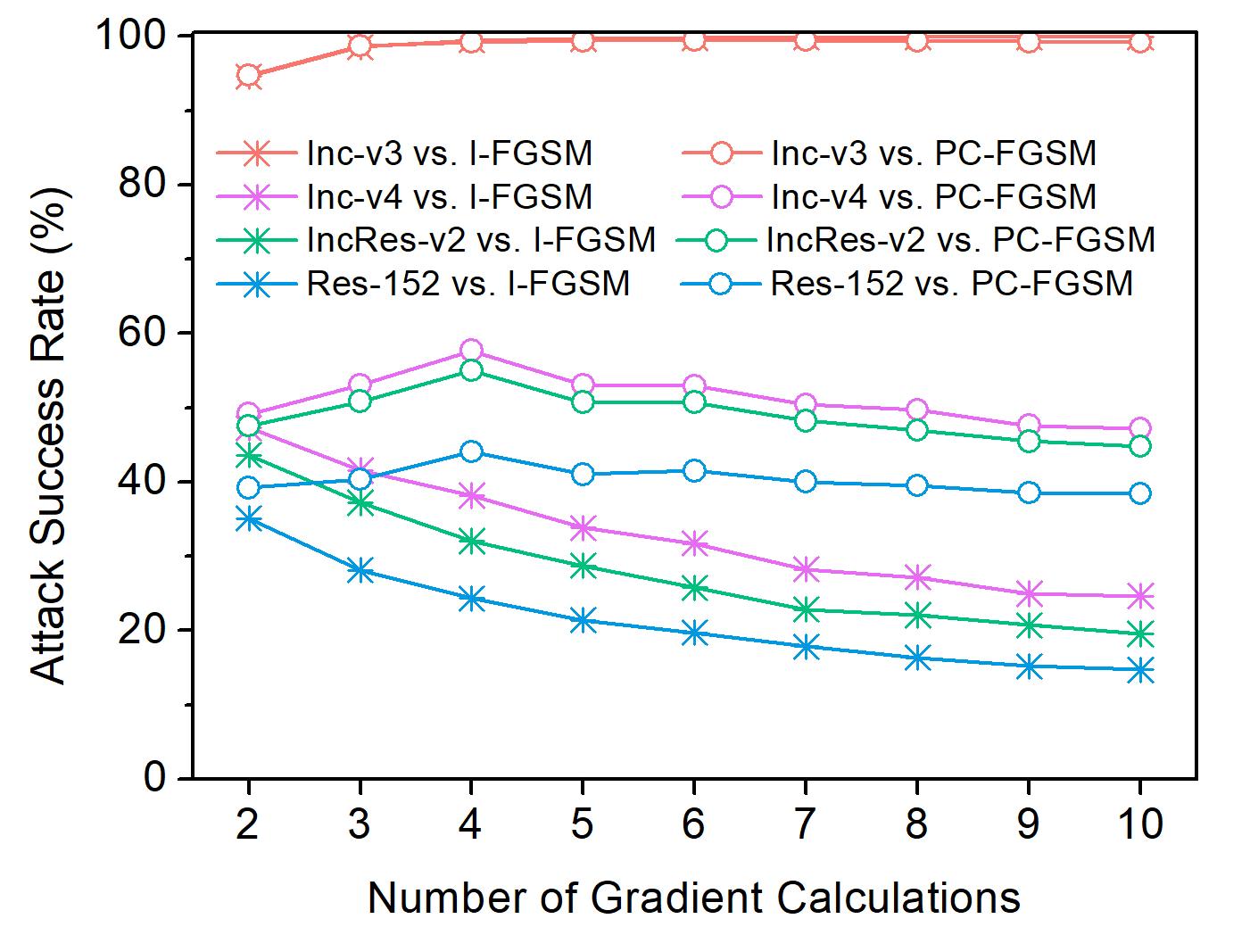}
        \caption{}
        \label{subfigure 4.1}
    \end{subfigure}
     \hfill
   \begin{subfigure}[b]{0.22\textwidth}
        \centering
       \includegraphics[width=1\textwidth]{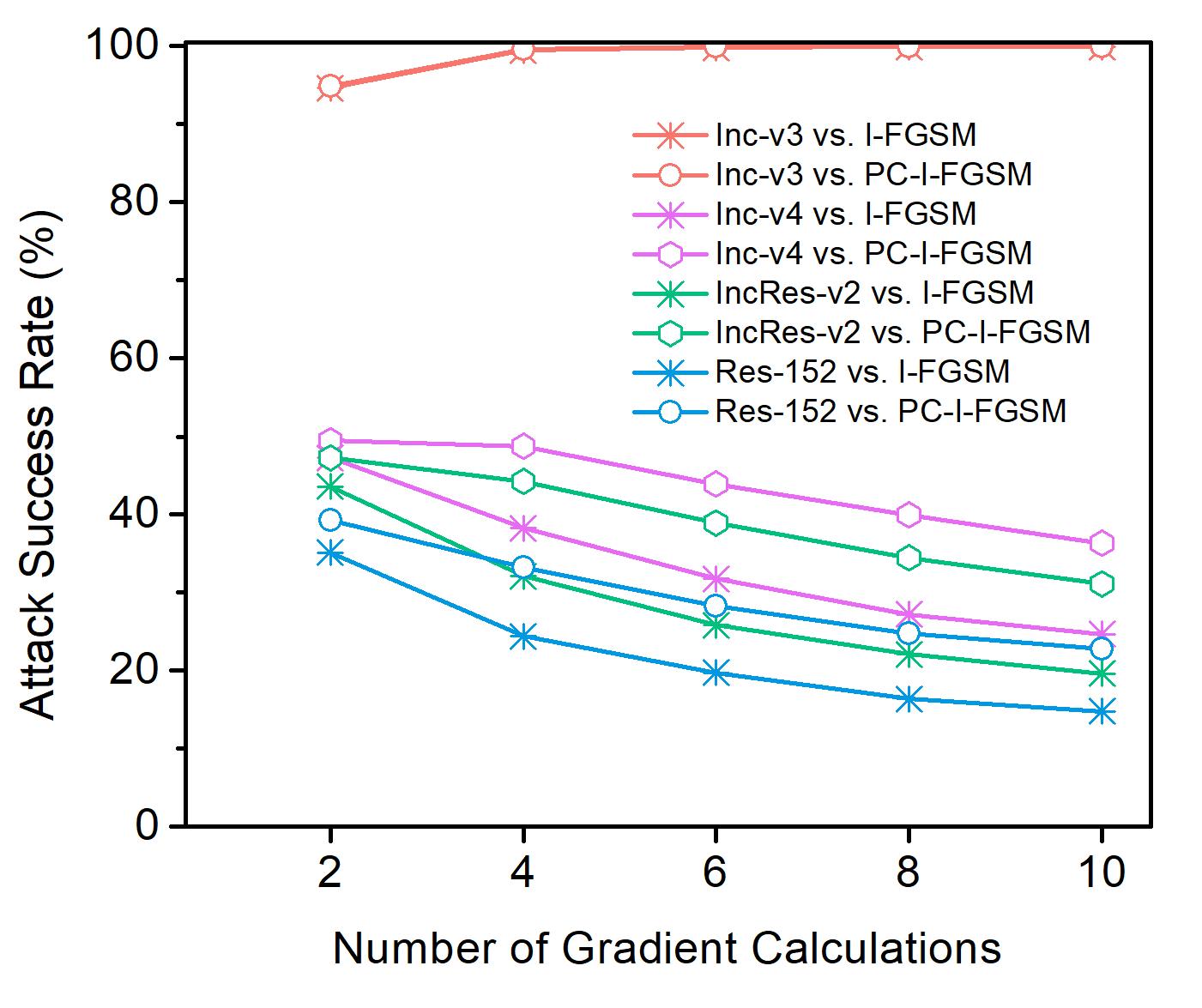}
        \caption{}
        \label{subfigure 4.2}
    \end{subfigure}
    \vspace{-0.3cm}
\caption{The attack success rates (\%) of our proposed PC-based attacks and the baseline attacks with the same number of iterations. (a) Comparison of I-FGSM and PC-FGSM; (b) Comparison of I-FGSM and PC-I-FGSM.}
\label{Fig:4} 
\vspace{-0.4cm}  
\end{figure}

Firstly, our proposed PC-FGSM and the three baseline attacks are compared.
For our proposed PC-FGSM, the number of predictions $K$ is selected in the range of 1 to 9, which corresponds to the number of gradient calculations from 2 to 10.
For the three baseline attacks, the number of iterations $T$ is selected in the range of 2 to 10, which also corresponds to the number of gradient calculations from 2 to 10.
We utilize PC-FGSM, I-FGSM, MI-FGSM, and NI-FGSM to attack Inc-v3 to obtain adversarial examples, and the obtained adversarial examples are tested on four normally trained models (\textit{i.e.}, Inc-v3, Inc-v4, IncRes-v2, and Res-152). 
The attack success rates of PC-FGSM and I-FGSM are shown in Figure \ref{Fig:4}\textcolor{red}{(a)}, and the experimental results of PC-FGSM and other two baseline attacks (\textit{i.e.}, MI-FGSM and NI-FGSM) are shown in Appendix C. 
The label “model vs. method” in Figure \ref{Fig:4} has a similar meaning as the label in Figure \ref{Fig:3}.
As seen, with the same number of gradient calculations, PC-FGSM can generally achieve higher attack success rates than I-FGSM. When the number of gradient calculations is small, PC-FGSM can also achieve a higher attack success rate than MI-FGSM and NI-FGSM in general.

Next, the attack success rates of our proposed PC-I-FGSM, PC-MI-FGSM, and PC-NI-FGSM are compared with that of the three baseline attacks (\textit{i.e.}, I-FGSM, MI-FGSM, and NI-FGSM). 
In order to ensure that the comparison is performed with the same number of gradient calculations, the number of predictions $K=1$ and iterations $T=1, 2, 3, 4, 5 $ are selected in our proposed PC-based attacks, and for the baseline attacks, the number of iterations is selected as $T=2, 4, 6, 8, 10$. The above-mentioned six methods are used to attack Inc-v3 to generate adversarial examples, and the obtained adversarial examples are tested on four normally trained models.
The experimental results of I-FGSM and PC-I-FGSM are shown in Figure \ref{Fig:4}\textcolor{red}{(b)}, and the other two comparisons (\textit{i.e.}, the comparison between MI-FGSM and PC-MI-FGSM, and the comparison between NI-FGSM and PC-NI-FGSM) are provided in Appendix C.
It can be observed that with the same number of gradient calculations, the attack success rates of our proposed PC-based attacks (\textit{i.e.}, PC-I-FGSM, PC-MI-FGSM, and PC-NI-FGSM) are generally higher than those of the baseline attacks.

\section{Conclusion}
In this paper, we first uncover the close relationship between the gradient-based attacks and the numerical methods for solving ODE. Based on this, a series of new PC-based attacks are proposed. The main advantages of the proposed PC-based attacks are as follows. 1) The adversarial examples generated by PC-based attacks can attack the white-box models more efficiently, meanwhile possess higher transferability against the black-box models and defense models. 2) The proposed PC-based attacks exhibit good extensibility and can be applied to almost all gradient-based attacks easily. 3) The PC-based attacks have high execution efficiency, \textit{i.e.}, with the same number of iterations (or the same number of gradient calculations), the proposed PC-based attacks can achieve higher attack success rates in general.

{\small
\bibliographystyle{ieee_fullname}
\bibliography{ref}
}

\end{document}